\documentclass{jpp}
\usepackage{graphicx}
\usepackage{epstopdf, epsfig, amsopn, amsmath}
\usepackage{color}
\usepackage{mathtools}
\usepackage{booktabs}

\newcommand{\bs}[1]{\boldsymbol #1}
\newcommand{\ee}{\mathrm{e}}
\newcommand{\dif}{\mathrm{d}}
\newcommand{\ii}{\mathrm{i}}
\DeclareMathOperator{\sinc}{sinc}

\shorttitle{Spectral method for non-periodic gyroaveraged electrostatic potential}
\shortauthor{J. Guadagni and A. J. Cerfon}

\title{Fast and spectrally accurate evaluation of gyroaverages in non-periodic gyrokinetic-Poisson simulations}

\author{J. Guadagni and
  A. J. Cerfon\corresp{\email{cerfon@cims.nyu.edu}}}

\affiliation{Courant Institute of Mathematical Sciences, New York University,
New York, NY 10012, USA}

\begin{document}

\maketitle

\begin{abstract}
We present a fast and spectrally accurate numerical scheme for the evaluation of the gyroaveraged electrostatic potential in non-periodic gyrokinetic-Poisson simulations. Our method relies on a reformulation of the gyrokinetic-Poisson system in which the gyroaverage in Poisson's equation is computed for the compactly supported charge density instead of the non-periodic, non-compactly supported potential itself. We calculate this gyroaverage with a combination of two Fourier transforms and a Hankel transform, which has the near optimal run time complexity $O(N_{\rho}(P+\hat{P})\log(P+\hat{P}))$, where $P$ is the number of spatial grid points, $\hat{P}$ the number of grid points in Fourier space, and $N_{\rho}$ the number of grid points in velocity space. We present numerical examples illustrating the performance of our code and demonstrating geometric convergence of the error.
\end{abstract}

\section{Introduction}
Plasmas in which the characteristic collision time is long compared to the time scales of the processes of interest must usually be described with kinetic equations. In general, phase space is 6-dimensional, so solving such equations numerically is computationally costly. For phenomena that are slow compared to the period of gyromotion of the particles, significant reduction in computational time and complexity can be obtained by averaging the equations over the gyromotion -- an operation known as gyroaveraing. The formal theory for this averaging procedure is called gyrokinetic theory \citep{Rutherford68,Taylor68,Catto,Catto_true,Antonsen,Frieman,Brizard}. It has been widely and successfully implemented in a large number of numerical codes, mostly for the study of magnetic confinement fusion and astrophysical plasmas \citep{Candy,Garbet10,Numata10,Gorler}. Many recent discoveries in plasma physics relied heavily on gyrokinetic simulations produced by these codes, particularly for problems in which kinetic turbulence plays a central role.

Despite the remarkable achievements of the solvers mentioned above, there remain questions regarding the numerical implementation of the gyrokinetic equations that are not entirely resolved yet. In this article, we address one of them, namely the fast and high-order accurate numerical evaluation of gyroaveraged quantities in settings in which the periodicity of the physical quantities cannot be assumed. The issue can be summarized as follows. It is well-known that in Fourier space the gyroaveraging operation reduces to a multiplication by the Bessel function $J_{0}$. This fact is conveniently used by the solvers which assume periodicity of the physical quantities, called local codes, turning gyroaveraging into a fast and high order accurate operation. It is, however, not straightforward to use this fact in non-periodic settings due to difficulties associated with the evaluation of the Fourier transform in these situations \citep{Crouseilles, Steiner}. Consequently, two alternative approaches have been followed. A first approach is to replace the Bessel function with a Pad\'e expansion approximation. The associated expansion for the gyroaveraging operation in Fourier space can then be transformed back to real space, and the gyroaverage is the solution of a tractable partial differential equation \citep{Sarazin,Steiner}. The Pad\'e approximation approach has the advantage of being fast, but is well-known to cause an overdamping of small scales, which limits its accuracy \citep{Steiner}. A more common approach is to evaluate directly the gyroaverage integral through numerical quadrature, relying on interpolation on points along the gyroring of the function which is gyroaveraged \citep{Jolliet,Crouseilles,Gorler,Steiner}. By choosing high order interpolation schemes, high order accuracy can be achieved with this method, but with a larger computational cost than with a Pad\'e based approach \citep{Steiner}.

In this article, we present a different strategy to calculate the gyroaveraged electrostatic potential, which leads to a spectrally convergent numerical scheme and a nearly optimal computational complexity, namely $O (N_{\rho}(P+\hat{P})\log (P+\hat{P}))$, where $P$ is the number of grid points in the spatial domain, $\hat{P}$ the number of grid points in Fourier space, and $N_{\rho}$ the number of grid points in velocity space. Our approach relies on a reformulation of the gyrokinetic-Poisson system in which Poisson's equation is not solved for the electrostatic potential $\Phi$ but instead for the gyroaverage of $\Phi$ at fixed guiding center position $\bs R$, which we call $\chi$. In that modified gyrokinetic-Poisson system, the only gyroaverage which needs to be evaluated numerically is the gyroaverage of the charge density, which is a compactly supported function. We are thus able to compute the Fourier transform of the charge density, and rely on the standard multiplication by the Bessel function $J_{0}$ in Fourier space and the Hankel transform to evaluate its gyroaverage. At that point, the desired spectral convergence and optimal run time complexity follow immediately from the adoption of well-known spectrally accurate algorithms with nearly optimal complexity for the calculation of the forward and inverse Fourier transforms and of the Hankel transform. 

The structure of the article is as follows. In section \ref{sec:formulation}, we present the gyrokinetic-Poisson system we are interested in and derive our reformulation of the system of equations, which allows the use of a Fourier representation for the calculation of the gyroaverage even for non-periodic settings. In section \ref{scheme}, we describe our numerical scheme by decomposing the gyroaverage operation into more fundamental mathematical operations. Note that our scheme relies on the sequential implementation of very well-known algorithms, which are open-source and readily available for implementation by any user. In section \ref{results}, we focus on functions which can be gyroaveraged analytically and compare the analytic results with our numerical results to demonstrate the spectral convergence of the numerical error. We summarize our work in section \ref{conclusion} and suggest directions for future work.
 
\section{New formulation for the gyrokinetic-Poisson equations}\label{sec:formulation}

In this section, we introduce the gyrokinetic-Poisson system of equations we will consider in the remainder of this article, and derive a reformulation of the system which is well-suited for the new numerical scheme for gyroaveraging we propose. We chose these gyrokinetic-Poisson equations for two reasons. First, their relative simplicity allows us to focus on the mathematical aspects of our numerical method for gyroaveraging, which is the main point of the article. Second, the equations correspond to a good model for the study of intense non-neutral beams in cyclotrons and vacuum tubes with high magnetic fields.

\subsection{Avoiding gyroaverages of the electrostatic potential}
We consider the simple situation of a two-dimensional single-species plasma in the $xy$-plane immersed in a constant and uniform magnetic field $\bs B=B_{0}\bs e_{z}$. We define the gyrofrequency $\Omega=qB_{0}/m$, with $q$ the charge of the particles in the plasma, and $m$ their mass. The position of a given particle is $\bs r=\bs R + \bs\rho$, where $\bs R$ is the guiding centre (or gyrocentre) position, and $\boldsymbol{\rho}$ is the Larmor radius vector $\boldsymbol{\rho}=-\rho\sin\gamma\bs e_{x}+\rho\cos\gamma\bs e_{y}$, where $\gamma$ is the gyroangle, and $(\bs e_{x},\bs e_{y},\bs e_{z})$ is the standard Cartesian orthonormal basis of $\mathbb{R}^3$, with $\bs e_{z}$ aligned with the magnetic field. If velocities are normalised to the thermal speed $v_{\textrm{th}}$ and  spatial scales are normalised to the Larmor radius $\rho_{L}=v_{\textrm{th}}/\Omega$, the gyrokinetic-Poisson equations for the gyrocentre distribution function $f(\bs R,\rho,t)$ are given by \citep{Hazeltine,Plunk}
\begin{gather}
\frac{\partial f}{\partial t}+\bs e_{z}\times\langle\bs\nabla_{\bs r}\Phi\rangle_{\bs R}\bs\cdot\bs\nabla_{\bs R} f=0\label{fullfgyro}\\
\nabla_{\bs r}^2\Phi(\bs r,t) =-\int_{0}^{+\infty}\int_{0}^{2\pi}f(x+\rho\sin\gamma,y-\rho\cos\gamma,\rho,t)\,\rho\,\dif\rho \dif\gamma\equiv -2\upi \tilde{f}(x,y,t)\;,\; \label{gyroPoisson}
\end{gather}
where
\begin{equation}
\tilde{f}(\bs r,t)=\frac{1}{2\upi}\int_{0}^{+\infty}\int_{0}^{2\upi}f(x+\rho\sin\gamma,y-\rho\cos\gamma,\rho,t)\,\rho\,\dif\rho \dif\gamma \;.\label{totalgyro}
\end{equation}
In equation (\ref{fullfgyro}), $\langle\cdot\rangle_{\mathbf{R}}$ represents the gyroaverage at fixed guiding centre position $\mathbf{R}=X\bs e_{x}+Y\bs e_{y}$:
\begin{equation}
\langle\Phi\rangle_{\bs R}=\frac{1}{2\upi}\int_{0}^{2\upi}\Phi(X-\rho\sin\gamma,Y+\rho\cos\gamma,t)\,\dif\gamma\;,\;\label{potgyroaveraged}
\end{equation}
where $X$ and $Y$ are held fixed. We chose the relatively simple gyrokinetic system of equations (\ref{fullfgyro}) and (\ref{gyroPoisson}) to better focus on the central question of this paper, namely the fast and accurate evaluation of the averages over the gyroangle $\gamma$ appearing in both equations. However, the method we present here is applicable to the more general gyrokinetic systems commonly used to study astrophysical and fusion plasmas. We should mention that equations (\ref{fullfgyro}) and (\ref{gyroPoisson}) are a surprisingly accurate description of the dynamics of a beam of charged particles in the plane perpendicular to the magnetic field in high intensity cyclotrons \citep{CerfonPRSTAB,GuadagniThesis,CerfonPRL}. With this application in mind, we want to allow boundary conditions on $\Phi$ that are not periodic, such as free space boundary conditions for instance.

If one has a numerical method to accurately evaluate $\langle\bs\nabla_{\bs r}\Phi\rangle_{\bs R}$ on the desired numerical grid, then a number of established numerical schemes are available to advance $f$ in time according to equation (\ref{fullfgyro}) \citep{Peterson,GuadagniThesis}. Clearly, the challenge that is specific to gyrokinetics is the numerical evaluation of the gyroaverage for the charge density in equation (\ref{gyroPoisson}) and of the gyroaveraged potential (\ref{potgyroaveraged}) when these quantities are not periodic. In the introduction, we have mentioned popular methods to accomplish this task. We propose a different approach, based on Fourier transforms, which leads to high-order accurate answers. Such an approach is not practical in the current formulation of the problem since $\Phi$ is not periodic and is unbounded for free space boundary conditions. Our first step therefore consists of casting equations (\ref{fullfgyro}) and (\ref{gyroPoisson}) in a form which is compatible with a Fourier representation.

For our simple geometry, $\nabla_{\bs r}=\nabla_{\bs R}$, so it is straightforward to re-express (\ref{fullfgyro}) and (\ref{gyroPoisson}) in terms of equations for quantities which only depend on the guiding centre position $(X,Y)$:
\begin{gather}
\frac{\partial f}{\partial t}+\bs e_{z}\times\bs\nabla_{\bs R}\langle\Phi\rangle_{\bs R}\bs\cdot\bs\nabla_{\bs R} f=0\label{fullfgyro2}\\
\nabla_{\bs R}^2\Phi(\bs R,t) = -2\upi \tilde{f}(X,Y,t)\;,\;\label{gyroPoisson2}
\end{gather}
where we also made use of the fact that the gradient operator commutes with the gyroaveraging operator \citep{GuadagniThesis}. Since $\Phi$ is not periodic and unbounded, it does not have a Fourier transform. Computing $\langle\Phi\rangle_{\bs R}$ using standard Fourier techniques is therefore not an available option. The idea instead is to define a new potential-like function $\chi(\bs R,\rho,t)=\langle\Phi\rangle_{\bs R}(\bs R,\rho,t)$ which we call the gyropotential. The key then is to not evaluate $\chi$ as given by its definition, but instead to see that it is the solution of the Poisson equation
\begin{equation}
\nabla^2\chi = \frac{1}{2\upi}\int_{0}^{2\upi}\tilde{f}(X+\rho\sin\gamma,Y-\rho\cos\gamma,t)\,\dif\gamma\equiv -2\upi\langle\tilde{f}\rangle_{\bs R}\;,
\end{equation}
which we obtained by gyroavering (\ref{gyroPoisson2}) holding the guiding centre position $\bs R$ fixed, and using once more the fact that the gradient operator commutes with the gyroaveraging operator. At this point, we have turned (\ref{fullfgyro}) and (\ref{gyroPoisson}) into the following new system of equations:
\begin{gather}
\frac{\partial f}{\partial t}+\bs e_{z}\times\bs\nabla\chi\bs\cdot\bs\nabla f=0\label{fullfgyro3}\\
\nabla^2\chi = -2\upi\langle\tilde{f}\rangle_{\bs R}\;. \label{gyroPoisson3}
\end{gather}
Unsurprisingly, this system shares many similarities with the two-dimensional inviscid Euler equations in vorticity-streamfunction form \citep{Plunk,CerfonPRL}. The major difference with the Euler equations is that the term $\langle\tilde{f}\rangle_{\bs R}$ couples the dynamics at different values of $\rho$. In the context of the present article, the significant aspect of the system of equations above is that it has the desirable property of only involving gyroaverages of $\tilde{f}$, which, unlike $\Phi$, can be approximated numerically by a compactly supported function. Once $\langle\tilde{f}\rangle_{\bs R}$ is known, Poisson's equation (\ref{gyroPoisson3}) can be solved with standard methods. 

It may first seem as if our reformulation of the gyrokinetic Vlasov-Poisson system has a large computational cost because $\chi$ is a function of the variable $\rho$. That means that Poisson's equation (\ref{gyroPoisson3}) has to be solved as many times as there are discretisation points for the $\rho$ variable. In practice, however, this is not a salient issue, for three reasons. First, there exist a wide choice of fast and high-order accurate Poisson solvers with nearly optimal computational complexity. For example, for the free space boundary conditions which are the relevant conditions for beam dynamics in cyclotrons, we may mention the solvers by Jiang et al. \citep{Jiang} and by Vico et al. \citep{Vico}, which solve Poisson's equation on a regular grid in $O(P\log P)$ time, where $P$ is the number of space discretisation points. Second, it is quite common in gyrokinetic simulations to have a very small number ($\leq 20$) of grid points for the $\rho$ variable \citep{Wilkening, CandyPrivate}. Lastly, the Poisson solve is rarely the rate limiting step in gyrokinetic simulations. This is particularly true for Particle-In-Cell (PIC) solvers, in which the particle operations typically dominate the computation and storage requirements \citep{Ricketson}.

\subsection{Limitations of a Fourier series expansion}

\begin{figure}
\centering\includegraphics[width=.5\linewidth]{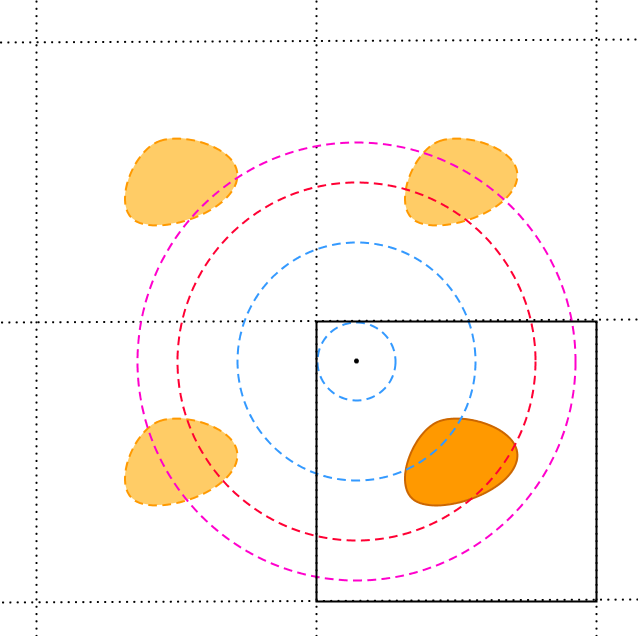}
\caption{Illustration of the difficulties associated with using a Fourier series representation for computing the gyroaverage of a distribution function $f$ is a non-periodic setting. The computational domain is the box bounded by the continuous lines. The boxes bounded by dashed lines represent some of the boxes induced by the periodicity of the Fourier series representation. The gyroaverage along the blue orbits yield correct results, while the gyroaverage along the red and purple orbits lead to fictitious contributions to the true gyroaverage.}
\label{fig:encroachment}
\end{figure}

Thus far, we have expressed the gyrokinetic-Poisson system of equations (\ref{fullfgyro}) -- (\ref{gyroPoisson}) in a form that only requires gyroaveraging a function which can be well approximated by a compactly supported function, given by the system of equations (\ref{fullfgyro3}) -- (\ref{gyroPoisson3}). This allows us to use a Fourier basis to represent that function and to calculate its gyroaverage in Fourier space. However, for problems which are not periodic, despite the compact support of $\tilde{f}$, a gyroaveraging scheme based on a Fourier series expansion for $\tilde{f}$ can only lead to highly accurate answers under specific conditions. To understand this point, consider the model charge density shown in Figure \ref{fig:encroachment}. The computational domain is represented by the black box in the lower right corner, and the support of the charge density is shown in orange there. By expanding $\tilde{f}$ in a Fourier basis, one would make the problem periodic, which is represented by the three boxes which are contiguous to the computational domain in Figure \ref{fig:encroachment} and represented with dashed lines.
Now, consider the four gyroaverages corresponding to different values of $\rho$, with common centre the black dot in the upper left corner of the computational domain, and shown in the figure as four dashed circles. The two inner circles, highlighted in blue, give the correct result for the gyroaverage: the average only involves contributions from the true charge density -- and no contribution at all for the innermost circle, as it should. However, the two outer circles, highlighted in red, give a false result for the gyroaverage. Along the third circle, the contribution from the true charge density is summed, but so are the contributions from the ``ghost", unphysical charge densities in the contiguous cells. For the outermost circle, corresponding to large $\rho$, the average should be zero, but it is not because the ``ghost" charge densities contribute to the average. 

A possible way to address this issue is to increase the size of the computational domain by padding it with zeroes in such a way that even for the maximum $\rho$ considered in the simulations, gyrocircles only average the true charge density. This idea is very similar to what is sometimes done to solve Poisson's equation with a plane wave representation \citep{Genovese}.  It has the clear disadvantage of leading to large computational domains, and therefore significantly more expensive evaluations of the Fast Fourier Transform (FFT) typically used to calculate the coefficients in the Fourier series. 

Now, while a Fourier series expansion leads to the complications discussed above, because it intrisically implies periodicity for the problem, an expansion on a continuous Fourier basis, i.e. through the Fourier transform, has all the desirable properties for a high-order accurate numerical scheme for gyroaveraging. This is what we discuss in the next section.

\subsection{Fourier and Hankel transform representation}
Consider a function $u(\bs x,\rho)$ on $\mathbb{R}^2\times\mathbb{R}_{\geq 0}$. The Fourier transform of $u$ with respect to the first two inputs is defined by
\begin{equation}
\hat{u}(\boldsymbol{\xi},\rho)\equiv(\mathcal{F}_{\bs x}u)(\boldsymbol{\xi},\rho)=\int_{\mathbb{R}^2}u(\bs x,\rho)\ee^{-\ii\boldsymbol{\xi}\bs\cdot\bs x}\,\dif\bs x\;.
\end{equation}
The inverse Fourier transform is defined by
\begin{equation}
u(\bs x,\rho)\equiv(\mathcal{F}^{-1}_{\boldsymbol{\xi}}\hat{u})(\bs x,\rho)=\frac{1}{4\upi^2}\int_{\mathbb{R}^2}\hat{u}(\boldsymbol{\xi},\rho)\ee^{\ii\boldsymbol{\xi}\bs\cdot\bs x}\,\dif\bs x\;.
\end{equation}
Finally, given a real-valued function $u(s)$ defined on $\mathbb{R}_{\geq 0}$, its Hankel transform of order 0 is defined by

\begin{equation}
(\mathcal{H}_{0}u)(s)=\int_{0}^\infty u(\rho)J_{0}(\rho s)\,\rho\,\dif\rho\;.
\label{eq:Hankel}
\end{equation}
where $J_{0}$ is the Bessel function of the first kind and of order 0. Introducing the gyroaverage at fixed particle position $\bs r=x\bs e_{x}+y\bs e_{y}$,
\begin{displaymath}
\langle f\rangle(x,y,\rho,t)\equiv\frac{1}{2\pi}\int_{0}^{2\pi}f(x+\rho\sin\gamma,y-\rho\cos\gamma,\rho,t) \dif\gamma\;,
\end{displaymath}
it is well-known that $\mathcal{F}\langle f\rangle(\boldsymbol{\xi},\rho,t)=J_{0}(\rho\xi)(\mathcal{F}f)(\boldsymbol{\xi},\rho,t)$. Therefore, we can write
\begin{equation}\label{totalgyro_1}
(\mathcal{F}\tilde{f})(\boldsymbol{\xi},t)=\int_{0}^{\infty}(\mathcal{F}\langle f\rangle)(\boldsymbol{\xi},\rho,t)\,\rho\,\dif\rho=\int_{0}^{\infty}J_{0}(\rho\xi)(\mathcal{F}f)(\boldsymbol{\xi},\rho,t)\,\rho\,\dif\rho=(\mathcal{H}_{0}\mathcal{F}f)(\boldsymbol{\xi},t)\;.
\end{equation}
And since we also have the identity
\begin{equation}\label{totalgyro_2}
(\mathcal{F}\langle\tilde{f}\rangle_{\bs R})(\boldsymbol{\xi},\rho,t)=J_{0}(\rho\xi)(\mathcal{F}\tilde{f})(\boldsymbol{\xi},t)=J_{0}(\rho\xi)(\mathcal{H}_{0}\mathcal{F}f)(\boldsymbol{\xi},t)\;,
\end{equation}
we obtain the desired expression for the only term which needs to be gyroaveraged in the system of equations (\ref{fullfgyro3}) - (\ref{gyroPoisson3}):

\begin{equation}\label{totalgyro_op}
\langle\tilde{f}\rangle_{\bs R} (\bs R,\rho,t)=\mathcal{F}^{-1}\left(J_{0}(\rho\xi)\mathcal{H}_{0}\mathcal{F}f\right)(\bs R,\rho,t)\equiv\mathcal{G}f(\bs R,\rho,t)\;.
\end{equation}
Note that there is a slight abuse of notation in equations (\ref{totalgyro_1})--(\ref{totalgyro_op}) above: in (\ref{eq:Hankel}) we have defined the Hankel transform for functions of a single variable, but $(\mathcal{F}f)$ clearly is not. 

Thus, to summarize section \ref{sec:formulation}, we have shown that the gyrokinetic-Poisson system can be rewritten as
\begin{gather}
\frac{\partial f}{\partial t}+\bs e_{z}\times\bs\nabla\chi\bs\cdot\bs\nabla f=0\label{fullfgyro4}\\
\nabla^2\chi = -2\upi\mathcal{G}f(\bs R,\rho,t)\;. \label{gyroPoisson4}
\end{gather}
All the gyroaveraging operations are contained in the operator $\mathcal{G}$ that operates on the function $f$, which is bounded and has compact support. Furthermore, if $f(X,Y,\rho,t)$ and $\chi(X,Y,\rho,t)$ are known at a given time $t$, the physical potential $\Phi(X,Y,t)$ and the number density $n(X,Y,t)$ are given by $\Phi(X,Y,t)=\chi(X,Y,0,t)$ and $n(X,Y,t)=2\pi\mathcal{G}f(X,Y,0,t)$. In other words, they do not require additional computation, and one just needs to read the values of $\chi(X,Y,\rho,t)$ and $2\pi\mathcal{G}f(X,Y,\rho,t)$ corresponding to $\rho=0$.

It remains to explain how we discretise $f$ and the operators on the right hand side of (\ref{totalgyro_op}) in order to evaluate $ \mathcal{G}f(\bs R,\rho,t)$ numerically to high accuracy. This is the purpose of the next section.

\section{Numerical scheme}\label{scheme}

In this section, we present the algorithmic details of our numerical scheme for computing $ \mathcal{G}f(\bs R,\rho,t)$, which is designed to lead to high order accuracy and near optimal run-time complexity. We give a brief justification for our choice of grids in section \ref{sec:grids}, after which the presentation follows the natural decomposition of $\mathcal{G}$ into its elementary operators: we describe our scheme for calculating the Fourier transform $\mathcal{F}$ in Section \ref{sec:Fourier}, present our numerical method for computing the Hankel transform $\mathcal{H}_{0}$ in Section \ref{sec:Hankel}, and describe our scheme for the inverse Fourier transform $\mathcal{F}^{-1}$ in Section \ref{sec:FourierInverse}. We conclude Section \ref{scheme} by giving recommendations for the choice of grid sizes and resolutions in Section \ref{sec:GridSizes}.

\subsection{Grid choices}\label{sec:grids}
For the design of our numerical scheme, we chose to operate under the constraint that the spatial grid be uniformly spaced. The reason for this is that many of the popular and advanced schemes for both the Vlasov equation and Poisson's equation are either very difficult to implement on nonuniform grids, or simply do not work on such grids \citep{Peterson,Shu,Jiang,Vico}. As we will see in sections \ref{sec:Fourier} and \ref{sec:FourierInverse}, this constraint leads to computational costs which could have been avoided if we had given ourselves the freedom to use a non-equispaced grid for the spatial variables. Still, even with this constraint and its associated computational cost, the scheme we propose here can be categorised as a fast solver, in the sense that its run time complexity is $O(K\log K)$, where $K$ is the number of degrees of freedom in the problem.

In contrast to the spatial grid, we never solve any partial differential equation with respect to either the Fourier space coordinates $(\xi_{x},\xi_{y})$ or the $\rho$-coordinate. We take advantage of this fact by using Chebyshev grids for each, which give us access to highly efficient and accurate numerical quadrature schemes.

\subsection{The Fourier transform $\mathcal{F}$}\label{sec:Fourier}
We start by presenting our method for evaluating $\mathcal{F}f$. For the simplicity of the presentation, we focus on the case in which $f$ depends only on one spatial variable $x$. The method generalizes directly to the two-dimensional tensor grid which we use to compute $\mathcal{G}f$. Consider the function $u$ on $\mathbb{R}$. Its Fourier transform is
\begin{equation}
\hat{u}(\xi)=\int_{-\infty}^{\infty}u(x)\ee^{-\ii\xi x}\,\dif x\;.
\label{forwardFourierDef}
\end{equation}
We consider the case in which $u$ is compactly supported on the domain $I=[-\tfrac{a}{2},\tfrac{a}{2}]$, which is relevant to us, and write $u$ as the \textit{exact} series
\begin{gather}
u(x)=\left(\sum_{k=-\infty}^\infty
c_{k}\ee^{2\upi \ii kx/a}\right)\mathbf{1}_{I}(x)\label{eq:exact_sum}\\
c_{k}=\frac{1}{a}\int_{-a/2}^{a/2}u(x)\ee^{-2\upi \ii kx/a}\,\dif x\;.
\end{gather}
In (\ref{eq:exact_sum}), $\mathbf{1}_{I}$ is the indicator function for the interval $I$, defined by $\mathbf{1}_{I}(x)=1$ if $x\in I$, and $\mathbf{1}_{I}(x)=0$ otherwise. The Fourier transform of $u$ is then given by
\begin{equation}
\hat{u}(\xi)=a\sum_{k=-\infty}^{\infty}c_{k}\sinc\left(k-\frac{a\xi}{2\upi}\right)\;.
\label{forwardFourier}
\end{equation}
where $\sinc(z)\equiv\tfrac{\sin(\upi z)}{\upi z}$. We proceed as follows to compute the sum (\ref{forwardFourier}) in optimal time, with high accuracy, and for values $\xi$ on an arbitrary grid in the Fourier domain. The Fourier coefficients $c_{k}$ are calculated with a straight forward call to the FFT, after discretising $u$ on a uniform grid. The complexity of this operation is $O(N\log N)$, where $N$ is the number of discretisation points for the $x$-grid. Once the $c_{k}$'s are known, we would like to evaluate the sum (\ref{forwardFourier}) on a $\xi$-grid which is optimal for the operations which will follow the calculation of $\mathcal{F}$ in the evaluation of the full operator $\mathcal{G}$. As mentioned in the previous section, that grid is a non-equispaced Chebyshev grid, which may in addition have different bounds than those naturally induced by the bounds of $I$ in the Fourier transform. To evaluate (\ref{forwardFourier}), we thus rely on the Fast Sinc Transform (FST) \citep{greengard1}, which is itself based on the Non-Uniform Fast Fourier Transform (NUFFT) \citep{dutt}, and which is freely available in a version described in \citep{greengard2,lee}. The computational complexity of the FST is $O((N+\hat{N})\log(N+\hat{N}))$, where $\hat{N}$ is the number of discretisation points for the $\xi$-grid. For the gyrokinetic-Poisson system (\ref{fullfgyro4})-(\ref{gyroPoisson4}) we are interested in, for which $f$ depends on two spatial variables, this becomes $O((P+\hat{P})\log(P+\hat{P}))$, where $P\sim N^2$ is the total number of spatial grid points, and $\hat{P}\sim\hat{N}^2$ is the total number of grid points in Fourier space. Since we need to repeat this operation for each value of $\rho$, the overall complexity of this step is $O(N_{\rho}(P+\hat{P})\log(P+\hat{P}))$, where $N_{\rho}$ is the number of grid points in velocity space.

Before closing this section, we mention an alternative way of evaluating (\ref{forwardFourierDef}) for a compactly supported function $u$. Since $u$ is compactly supported, it can be viewed as periodic on the domain of integration, so the trapezoidal rule provides a spectrally accurate scheme for the numerical evaluation of the integral \citep{TrefethenWeideman}. The discrete sum resulting from the application of the trapezoidal rule can then be computed with a direct application of the FFT \citep{patakigreengard}. The issue with this approach is that the number of grid points in real space determines the number of grid points in Fourier space. For most functions of physical interest, the representation in Fourier space is significantly more oscillatory than the representation in real space, so proper resolution in Fourier space requires significant oversampling in real space \citep{patakigreengard}, which is computationally costly. In contrast, the FST gives us the choice to have a larger number of grid points in Fourier space than the number of grid points we use in real space. Now, since the FST is more expensive than the regular FFT by a constant factor, it can be shown that the run time complexity of the two approaches is comparable. In our implementations of the scheme we present here, we favour the FST approach because we find the framework in which real space and Fourier space are decoupled elegant, convenient and efficient.

\subsection{The Hankel transform $\mathcal{H}_{0}$}\label{sec:Hankel}

The next step in the evaluation of $\mathcal{G}f$ corresponds to the calculation of the Hankel-like integral
\begin{equation}
g(\boldsymbol{\xi},t)=\int_{0}^\infty\hat{f}(\boldsymbol{\xi},\rho,t)J_{0}(\rho\xi)\,\rho\,\dif\rho\;,
\end{equation}
with $\xi=||\boldsymbol{\xi}||$. Since $\boldsymbol{\xi}$ and $t$ are fixed parameters in this integral, we simplify the notation and consider the computation of the integral
\begin{equation}
H(\xi)=\int_{0}^\infty h(\rho)J_{0}(\rho\xi)\,\rho\,\dif\rho\;.
\label{Hankelsimple}
\end{equation}
We consider in this work that to the desired numerical accuracy, $\hat{f}$ has compact support in the $\rho$-variable. In the context of Eq. (\ref{Hankelsimple}), this means that there exists a maximum $\overline{\rho}$ such that $h=0$ outside the interval $I_{\rho}
=[0,\overline{\rho}]$. To compute (\ref{Hankelsimple}) we discretise $I_{\rho}$ with a Chebyshev grid in $\rho$ and use Clenshaw-Curtis quadrature, which has geometric convergence for the class of functions we consider here \citep{trefethenClenshaw}:
\begin{equation}\label{CC}
H(\xi)\approx\sum_{k=1}^{N_{\rho}}w_{k}h(\rho_{k})J_{0}(\rho_{k}\xi)\rho_{k}\;,
\end{equation}
where $N_{\rho}$ is the number of Chebyshev grid points on the interval $I_{\rho}$, the $\rho_{k}$'s are the Chebyshev abcissae, and the $w_{k}$'s are the Clenshaw-Curtis quadrature weights. Now, observe that for fixed computational $\rho$- and $\xi$-grids, the values of $w_{k}J_{0}(\rho_{k}\xi)\rho_{k}$ can be precomputed for all $k$ and for all possible values of $\xi$. Hence, in practice, each individual Hankel integral can be computed as the inner product of one data vector which changes with each time step and one vector of fixed kernel weights. To be more precise, observe that in our case, $h$ is the function $\mathcal{F}f$, which does not only depend on $\rho$, but also on $\boldsymbol{\xi}$ and $t$, as can be seen in Equation (\ref{totalgyro_1}). Letting $h(\boldsymbol{\xi},\rho,t)=\mathcal{F}f(\boldsymbol{\xi},\rho,t)$ to simplify the notation in the matrices below, all Hankel integrals are computed at once by considering the $N_{\rho}\times\hat{P}$ matrices $\bs h$ and $\bs W$ defined by
\begin{displaymath}
\bs h=\begin{pmatrix}
h(\boldsymbol{\xi}_{1},\rho_{1},t) & h(\boldsymbol{\xi}_{2},\rho_{1},t)&\ldots& h(\boldsymbol{\xi}_{\hat{P}},\rho_{1},t)\\
h(\boldsymbol{\xi}_{1},\rho_{2},t) & h(\boldsymbol{\xi}_{2},\rho_{2},t)&\ldots&h(\boldsymbol{\xi}_{\hat{P}},\rho_{2},t)\\
\vdots & \vdots & \ldots & \vdots\\
h(\boldsymbol{\xi}_{1},\rho_{N_{\rho}},t)&h(\boldsymbol{\xi}_{2},\rho_{N_{\rho}},t)&\ldots & h(\boldsymbol{\xi}_{\hat{P}},\rho_{N_{\rho}},t)
\end{pmatrix}
\end{displaymath}
and
\begin{displaymath}
\bs W=\begin{pmatrix}
w_{1}J_{0}(\rho_{1}\xi_{1})\rho_{1} & w_{1}J_{0}(\rho_{1}\xi_{2})\rho_{1}&\ldots & w_{1}J_{0}(\rho_{1}\xi_{\hat{P}})\rho_{1}\\
w_{1}J_{0}(\rho_{2}\xi_{1})\rho_{2} & w_{1}J_{0}(\rho_{2}\xi_{2})\rho_{2}&\ldots & w_{1}J_{0}(\rho_{2}\xi_{\hat{P}})\rho_{2}\\
\vdots & \vdots & \ldots & \vdots\\
w_{1}J_{0}(\rho_{N_{\rho}}\xi_{1})\rho_{N_{\rho}}&w_{1}J_{0}(\rho_{N_{\rho}}\xi_{2})\rho_{N_{\rho}}&\ldots & w_{1}J_{0}(\rho_{N_{\rho}}\xi_{\hat{P}})\rho_{N_{\rho}}
\end{pmatrix}\;,
\end{displaymath}
where $\boldsymbol{\xi}_j$ is now viewed as the vector whose components are the coordinates of the $j$th pair of the $\hat{P}$ Fourier coordinate pairs. Note that $\bs W$ is fixed for all time and can be stored after its initial computation. The right-hand side of (\ref{totalgyro_1}), $\mathcal{H}_{0}\mathcal{F}f(\boldsymbol{\xi},t)$, is then calculated by computing the entrywise product of $\bs h$ and $\bs W$ and summing the columns of that matrix. In mathematical notation, we may write
\begin{equation}\label{eq:schur}
\mathcal{H}_{0}\mathcal{F}f(\boldsymbol{\xi},t)\approx \left(\sum_{i=1}^{N_{\rho}}(\bs h \odot \bs W)_{i1},\ldots,\sum_{i=1}^{N_{\rho}}(\bs h \odot \bs W)_{i\hat{P}}\right)\;,
\end{equation}
where $\odot$ represents the Schur product \citep{Davis}, i.e. entrywise product. The run-time complexity of this operation is $O(N_{\rho}\hat{P})$. As a result, when evaluating $\mathcal{G}f$ numerically with the method we present here, the total time spent computing Hankel transforms is negligible compared to the total time spent computing forward Fourier transforms.

\subsection{The inverse Fourier transform $\mathcal{F}^{-1}$}\label{sec:FourierInverse}
After a straightforward multiplication by $J_{0}(\rho\xi)$, all that is left to compute $\mathcal{G}f$ is the computation of the inverse Fourier transform $\mathcal{F}^{-1}$. As before, we focus here on the case of the one-dimensional inverse Fourier transform defined by
\begin{equation}
u(x)=\frac{1}{2\upi}\int_{-\infty}^{\infty}\hat{u}(\xi)\ee^{\ii\xi x}\,\dif\xi\;.
\label{InverseFourier}
\end{equation}
In principle, this computation can be done with the same numerical tools presented in Section \ref{sec:Fourier}. There are however two key points which we need to revisit. The first point to consider is that we relied on the fact that $u$ was compactly supported (to within the desired numerical accuracy) to reduce the Fourier transform integral to the finite interval $I$. From the Fourier uncertainty principle \citep{Hardy,HoganFourier}, it is not clear that the corresponding Fourier transform has numerical compact support on a finite interval of reasonable size. It may first seem as if this observation strongly limits the class of functions for which $\mathcal{G}f$ can be accurately computed with our numerical scheme. The functions for which both the Fourier transform and the inverse Fourier transform can be computed without significant loss of accuracy by restricting the quadratures to finite intervals with reasonable sizes are functions which can be represented as a Gaussian distribution function plus a small deviation from the Gaussian behavior \citep{Hardy,SteinShakarchi}. However, this turns out to be too pessimistic an estimate in practice. This is because our scheme does not require the computation of the inverse Fourier transform of $\mathcal{F}f$, but instead of $J_{0}(\rho\xi)\mathcal{H}_{0}\mathcal{F}f$. For finite $\rho$, $J_{0}(\rho\xi)$ is a decaying function of $\xi$, so the (numerical) compact support of $J_{0}(\rho\xi)\mathcal{H}_{0}\mathcal{F}f$ is contained within, and typically much smaller than, that of $\mathcal{F}f$. In Section \ref{sec:GridSizes} we provide guidelines for choosing the size $\hat{a}$ of the Fourier domain and the number $\hat{N}$ of the Chebyshev points which we have empirically found to provide accurate results in a robust manner for a wide class of distribution functions, including sub-Gaussian and super-Gaussian distribution functions. 

The second point to revisit concerns the numerical method to be used to evaluate (\ref{InverseFourier}). As we just mentioned, we could rely on the symmetry of the Fourier transform to reuse the methods employed in Section \ref{sec:Fourier} for the computation of (\ref{InverseFourier}). This is perfectly acceptable, but since one application of the FST requires the use of approximately four NUFFTs, for a given target accuracy this is not as efficient \citep{GuadagniThesis} as computing (\ref{InverseFourier}) directly with the NUFFT through the expression
\begin{equation}
u(x_{j})\approx\frac{1}{2\upi}\sum_{k=1}^{\hat{N}}w_{k}\hat{u}(\xi_{k})\ee^{\ii\xi_{k}x_{j}}\;,
\end{equation}
where the $w_{k}$ are Clenshaw-Curtis weights associated with the Chebyshev grid $\xi_{k}$ we use in Fourier space, the $x_{j}$ are the nodes of the real space equispaced grid. The run time complexity of the inverse Fourier transform computed in this manner is $O((N+\hat{N})\log(N+\hat{N}))$. So for a distribution functions which depend on two spatial variables, it is $O((P+\hat{P})\log(P+\hat{P}))$, with $P\sim N^2$ and $\hat{P}\sim \hat{N}^2$, and since this operation has to be done for every possible value of $N_{\rho}$, the overall run time complexity of this step is $O(N_{\rho}(P+\hat{P})\log(P+\hat{P})))$.

To complete the description of our algorithm, we need to specify how to choose $\hat{a}$ and $\hat{N}$, since the NUFFT gives us the freedom to choose them independently of $a$ and $N$. This is the purpose of the next section.

\subsection{Size of the computational domain and grid resolution}\label{sec:GridSizes}

In what follows, we assume that $\overline{\rho}$ is the maximum value of $\rho$ considered in the simulation. In other words, $f(x,y,\rho)$ has (numerical) compact support on $[-\tfrac{a}{2},\tfrac{a}{2}]^2\times[0,\overline{\rho}]$. 

\subsubsection{Size and resolution for the spatial domain}
Let $d_{\tilde{f}}$ denote the diameter of the smallest ball in $\mathbb{R}^2$ centered at the origin that contains the compact support of $\tilde{f}$. It can be shown that for fixed $\rho$, the support of $\mathcal{G}f(x,y,\rho)$ lies within the annular region $\sqrt{x^2+y^2}\in[\rho-\tfrac{1}{2}d_{\tilde{f}},\rho+\tfrac{1}{2}d_{\tilde{f}}]$ \citep{GuadagniThesis}. We thus take $a$ at least as large as $d_{\tilde{f}}+\overline{\rho}$. The quantity $d_{\tilde{f}}$ depends on time, so it must in principle be recomputed at every time step. However, in many situations of physical interest, $d_{\tilde{f}}$ as determined by the initial data $f(x,y,\rho,0)$ is a good enough guess for the entire simulation, and does not have to be adjusted at later times. This was for example the case for our gyrokinetic simulations of beam dynamics in cyclotrons, which are dominated by $\bs E\times\bs B$ physics, which led to differential rotation about the guiding magnetic field $\bs B$ but limited radial expansion of the beam distribution \citep{GuadagniThesis}.

The number $N$ of grid points in the $x$- and $y$-directions may be chosen arbitrarily, provided it is large enough to resolve the spatial variations of the distribution function one is interested in.

\subsubsection{Size and resolution for the Fourier domain}
We now provide simple guidelines for choosing the size $\hat{a}$ of each dimension of the Fourier domain, and the number of grid points in each dimension $\hat{N}$. These guidelines are based on elementary reasoning, which we have nonetheless found to be remarkably reliable for a wide range of distribution functions. Our heuristic argument is as follows. Consider the Gaussian distribution $u(x)=A\ee^{-bx^2}$, where $A$ and $b$ are constants. It has numerical support on the interval $I'=[-\frac{a'}{2},\frac{a'}{2}]$, where
\begin{equation}
a'=2\sqrt{-\frac{1}{b}\log\left(\frac{\epsilon^*}{A}\right)}\;.
\label{eqforaprime}
\end{equation}
The number $\epsilon^*$ is the absolute threshold that defines numerical compact support, and $\epsilon \equiv\tfrac{\epsilon^*}{A}$ is the relative threshold for compact support. For computations relying on double-precision floating-point format, one may for instance choose $\epsilon=10^{-16}$. The Fourier transform of $u$ is $\hat{u}(\xi)=A\sqrt{\frac{\upi}{b}}\ee^{-\xi^2/4b}$, which has numerical compact support on the interval $\hat{I}=\left[-\frac{\hat{a}}{2},\frac{\hat{a}}{2}\right]$ with $\hat{a}$ given by
\begin{equation}
\hat{a}=2\sqrt{-4b\log\left(\epsilon\sqrt{\frac{b}{\upi}}\right)}\;.
\label{eqforahat}
\end{equation}
Using (\ref{eqforaprime}) to express $b$ in terms of $a'$, we obtain the desired formula for the length of the compact support of $\hat{u}$ in terms of the length of the compact support for $u$:
\begin{equation}
\hat{a}(\epsilon,a')=\frac{8}{a'}\sqrt{\log\left(\frac{1}{\epsilon}\right)\log\left(\frac{a'}{2\epsilon}\sqrt{\frac{\upi}{\log\left(\frac{1}{\epsilon}\right)}}\right)}\;,
\label{ahatintermsofaprime}
\end{equation}
from which we can see that $\hat{a}\sim\tfrac{\sqrt{\log a'}}{a'}$. Although this formula holds exactly only for Gaussian distribution functions, we have empirically found it to be a reliable estimate for a wide range of functions. It is important here to note that in general $a'\neq a$, i.e. the length $a'$ for the compact support of $f$ may be different from the size of the spatial domain $a$. In principle $a'$ has to be computed at each time step, but in many situations of physical interest $a'$ may not vary much over the course of a simulation. As an illustration, as was the case for $d_{\tilde{f}}$, we found in our beam dynamics simulations that we were able to keep $a'$ fixed at its initial value determined by $f(x,y,\rho,0)$ and obtain accurate results. We observe finally that instead of using the formula given by (\ref{ahatintermsofaprime}) to estimate $\hat{a}$, a relatively straightforward (albeit possibly computationally costly) strategy, would be to calculate $\hat{a}$ numerically at each time step, or at regularly spaced time steps. We have not explored such a strategy in our simulations.

For the choice of the grid resolution in Fourier space, specified by $\hat{N}$, we also follow a simple reasoning. The general rule for the proper resolution of waves on a Chebyshev grid is to have at least $\upi$ Chebyshev nodes per wavelength on average \citep{trefethenspectral,marburg,trefethenapproximation}. If $k$ is the largest wavenumber associated with the oscillatory signal $\hat{u}$, then we take $\hat{N}=\frac{k\hat{a}}{2}$. Now, it is hard to predict how oscillatory $\hat{u}$ can be in the most general case. We consider the unfavorable scenario in which $u$ is the indicator function $u(x)=\mathbf{1}_{[-\frac{a'}{2},\frac{a'}{2}]}(x)$, with Fourier transform $\hat{u}(\xi)=a\sinc\left(\frac{a\xi}{2\pi}\right)$. This function is not smooth enough to be part of the class of functions for which our overall scheme for computing $\mathcal{G}f$ is reliable, but can serve as a good, often conservative estimate of how oscillatory $\hat{u}$ can be. For this function, we find $k\sim \tfrac{a'}{2}$. Now, accounting for the fact that in our simulations $u$ may not be symmetric with respect to the origin, and accounting for the oscillatory nature of the inverse Fourier kernel, we obtain the estimate $k\sim\frac{3}{2}a'$ for the maximum $k$ we may encounter \citep{GuadagniThesis}, so that $\hat{N}\sim \frac{3}{4}a'\hat{a}$.

\subsubsection{Resolution in $\rho$-space}
We have already explained that the domain for the $\rho$-variable is $[0,\overline{\rho}]$, where $\overline{\rho}$ is the smallest value such that $f(x,y,\rho)$ has numerical compact support for all values of $x$ and $y$ in $[-\tfrac{a}{2},\tfrac{a}{2}]^2$. The choice for the number of grid points $N_{\rho}$ to discretise the interval $[0,\overline{\rho}]$ depends on various considerations, some of which are not related to the computation of $\mathcal{G}f$. It is for example clear that $N_{\rho}$ should be large enough to properly resolve the details of the $\rho$-dependence of $f$. Even if so, we suggest here an initial guess for $N_{\rho}$ motivated by the need to accurately calculate $\mathcal{G}f$. Our reasoning is again quite elementary, since it does not take into account the oscillatory nature of $\tilde{f}(\xi_{x},\xi_{y},\rho,t)$ but the outcome has turned out to be a surprisingly robust guideline for choosing $N_{\rho}$. The Hankel kernel $J_{0}(\rho\xi)\rho$ is an oscillatory function with an approximate wavenumber of $k =\xi$, so the interval $[0, \overline{\rho}]$ has about $\frac{\overline{\rho}\xi}{2\upi}$ wavelengths. For $(\xi_{x},\xi_{y})\in [-\frac{\hat{a}}{2},\frac{\hat{a}}{2}]$, this quantity is maximized when $\xi=\tfrac{\hat{a}}{\sqrt{2}}$, which means that we need $N_{\rho}\sim\frac{\rho\hat{a}}{2\sqrt{2}}$, since we use a Chebyshev grid for the $\rho$-variable. 

\subsection{Summary}
\begin{figure}
\centering\includegraphics[width=.75\linewidth]{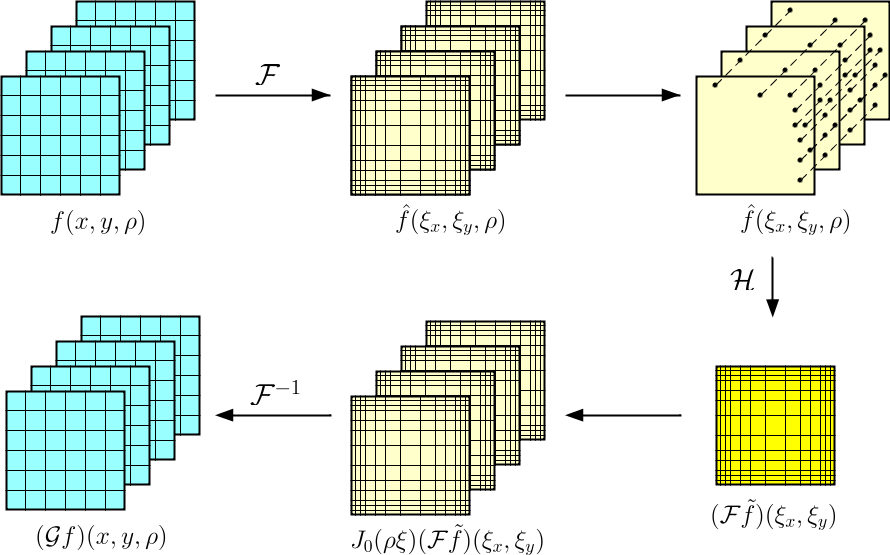}
\caption{Graphical representation of the successive steps involved in the numerical evaluation of $\mathcal{G}f$. The lines in each colored plane correspond to lines of constant $x$ and constant $y$, and we note the equispaced grid for the spatial variables, and the Chebyshev grid in Fourier space. The successive planes correspond to different values of the variable $\rho$. Since the computation of $\mathcal{G}f$ is done at a fixed time $t$, we omitted the time dependence in all the formulae in this figure.}
\label{fig:scheme}
\end{figure}

Our numerical scheme for computing $\mathcal{G}f$ is summarized in Figure \ref{fig:scheme}. In this figure, we omitted for simplicity the time dependence of all the quantities, since the computation of $\mathcal{G}f$ is done at fixed $t$. The parallel planes represent the spatial (in blue) or Fourier (in yellow) grids for different values of $\rho$. The Hankel step $\mathcal{H}$ collapses the $N_{\rho}$ planes into a single plane, which contains the data corresponding to the Fourier transform of $\tilde{f}$. Multiplying $\mathcal{F}\tilde{f}$ by $J_{0}(\rho\xi)$, we recreate $N_{\rho}$ panels on which we apply the inverse Fourier transform $\mathcal{F}^{-1}$ to obtain the desired $\mathcal{G}f$.

\section{Numerical examples}\label{results}
In this section we examine the accuracy and run time of our method by investigating a specific function $f(x,y,\rho)$ for which $\mathcal{G}f(x,y,\rho)$ can be computed analytically. Here, we have suppressed the dependence on $t$ for simplicity since the gyroaverages are computed once at each fixed time step. We consider the function
\begin{equation}\label{eq:exactf}
f(x,y,\rho) = \ee^{-A(x^2+y^2)}\ee^{-B\rho^2}\;,
\end{equation}
where $A$ and $B$ are positive constants. It can be shown that $\mathcal{G}f$ has the following closed-form formula:
\begin{equation} \label{eq:exactGf}
\mathcal{G}f(x,y,\rho) = \frac{1}{2(A+B)}\ee^{-\alpha(x^2+y^2+\rho^2)}I_0\left(2\alpha\rho\sqrt{x^2+y^2}\right)
\;,\end{equation}
where $\tfrac{1}{\alpha}=\tfrac{1}{A}+\tfrac{1}{B}$ and $I_0(z)$ is the modified Bessel function of the first kind of order 0. In Section \ref{sec:accuracy}, we focus on the convergence properties of our scheme, and in section \ref{sec:run_time} we present an analysis of the run times of our code for three levels of resolution: low, medium, and high.

\subsection{Accuracy of our scheme}\label{sec:accuracy}
For the sake of definiteness, in our experiments we put $A=B=15$. Consistent with
what we have already explained, we choose the following fixed interval sizes for the various grids involved in our numerical scheme.
\begin{align}
\mathrm{real\,\,\,space:}\quad& (x,y)\in [-3, 3] \\
\mathrm{Fourier\,\,\,space:}\quad&(\xi_x,\xi_y)\in[-66,66] \\
\mathrm{gyroradius:}\quad&\rho\in[0,1.55]\;.
\end{align}
We define the sampling rate of a particular grid to be the number of grid points per unit length. For a particular experiment, the sampling rates for the real space grid, Fourier space grid,
and gyroradius grid are denoted, respectively, $L_N$, $L_{\hat{N}}$, and $L_{N_{\rho}}$. In the language of the previous sections, it follows that $L_N=\tfrac{N}{a}$, $L_{\hat{N}}=\tfrac{\hat{N}}{\hat{a}}$, and $L_{N_{\rho}}=\tfrac{N_{\rho}}{\overline{\rho}}$.

\begin{figure}
\centering
\begin{minipage}[b]{0.48\linewidth}
\includegraphics[width=\linewidth]{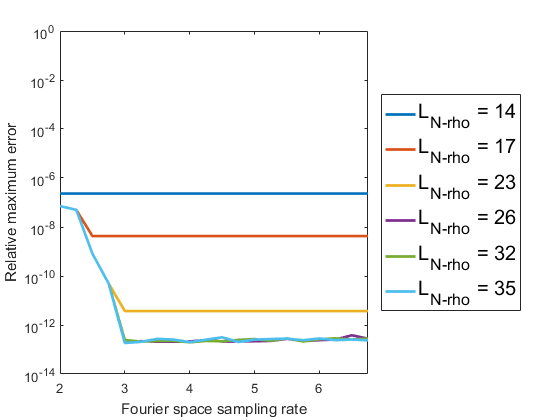}
\caption{Plot of $\epsilon_{\mathrm{abs}}$ as a function of $L_{\hat{N}}$ and selected values of $L_{N_{\rho}}$, for the fixed real sampling rate $L_N=12$.}
\label{fig:fourierfixedreal}
\end{minipage}
\hfill
\begin{minipage}[b]{0.48\linewidth}
\includegraphics[width=\linewidth]{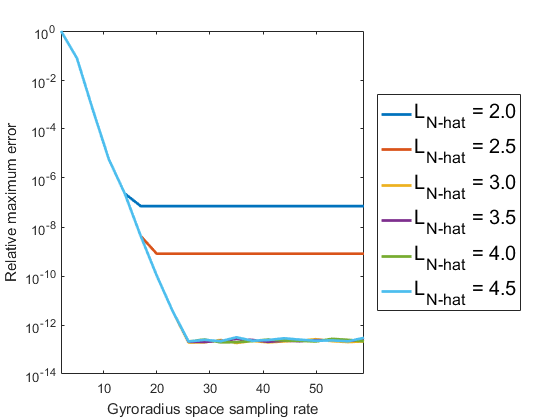}
\caption{Plot of $\epsilon_{\mathrm{abs}}$ as a function of $L_{N_{\rho}}$ and selected values of $L_{\hat{N}}$, for the fixed real sampling rate $L_N=12$.}
\label{fig:gyrofixedreal}
\end{minipage}
\begin{minipage}[b]{0.48\linewidth}
\includegraphics[width=\linewidth]{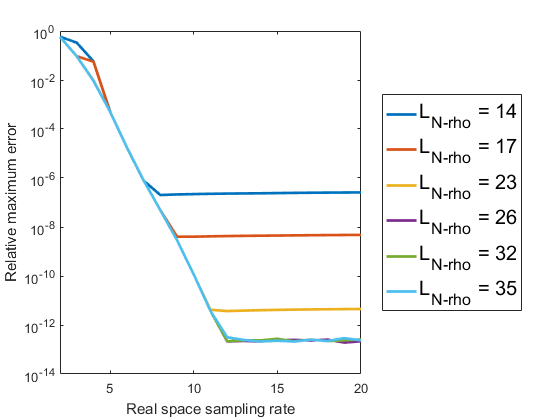}
\caption{Plot of $\epsilon_{\mathrm{abs}}$ as a function of $L_N$ and selected values of $L_{N_{\rho}}$, for the fixed Fourier sampling rate $L_{\hat{N}}=4.5$.}
\label{fig:realfixedfourier}
\end{minipage}
\hfill
\begin{minipage}[b]{0.48\linewidth}
\includegraphics[width=\linewidth]{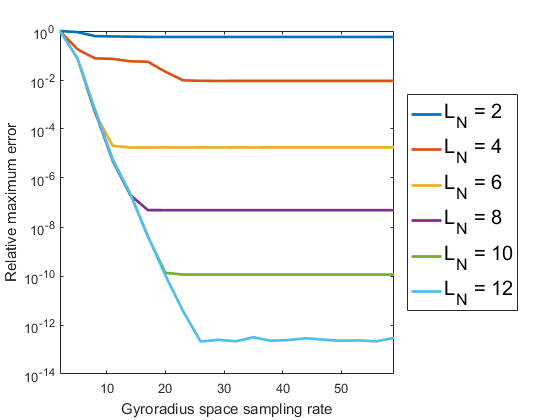}
\caption{Plot of $\epsilon_{\mathrm{abs}}$ as a function of $L_{N_{\rho}}$ and selected values of $L_N$, for the fixed Fourier sampling rate $L_{\hat{N}}=4.5$.}
\label{fig:gyrofixedfourier}
\end{minipage}
\begin{minipage}[b]{0.48\linewidth}
\includegraphics[width=\linewidth]{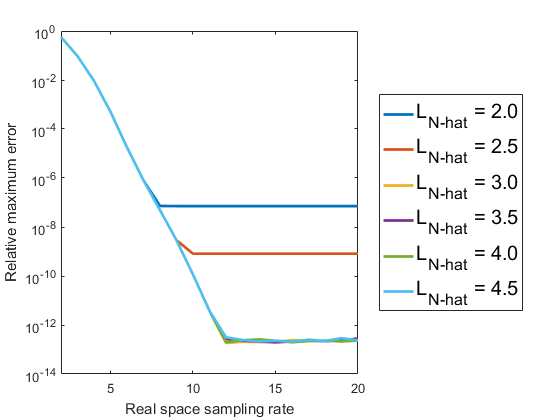}
\caption{Plot of $\epsilon_{\mathrm{abs}}$ as a function of $L_N$ and selected values of $L_{\hat{N}}$, for the fixed gyroradius sampling rate $L_{N_{\rho}}=35$.}
\label{fig:realfixedgyro}
\end{minipage}
\hfill
\begin{minipage}[b]{0.48\linewidth}
\includegraphics[width=\linewidth]{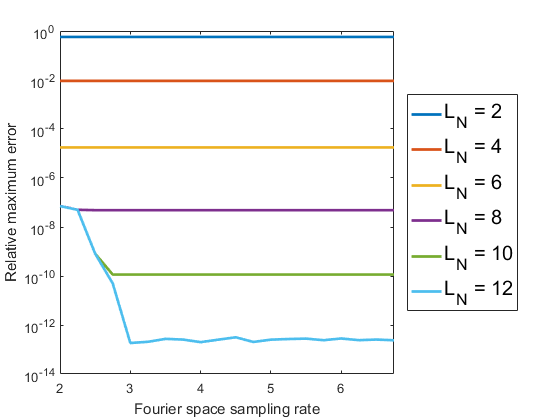}
\caption{Plot of $\epsilon_{\mathrm{abs}}$ as a function of $L_{\hat{N}}$ and selected values of $L_N$, for the fixed gyroradius sampling rate $L_{N_{\rho}}=35$.}
\label{fig:fourierfixedgyro}
\end{minipage}
\end{figure}

Let $\mathcal{G}f^{\mathrm{ex}}(x,y,\rho)$ denote the \textit{exact} function in (\ref{eq:exactGf}) and let $\mathcal{G}f^{\mathrm{num}}(x,y,\rho)$ denote the numerical approximation to $\mathcal{G}f^{\mathrm{ex}}$ obtained by applying our numerical scheme for gyroaveraging to the function $\tilde{f}$ in (\ref{eq:exactf}). For our purposes, we are concerned with the relative maximum error $\epsilon_{\mathrm{abs}}$, defined as
\begin{equation}
\epsilon_{\mathrm{abs}}=
\frac{\max_{i,j,k}\left|\mathcal{G}f^{\mathrm{ex}}(x_i,y_j,\rho_k)-\mathcal{G}f^{\mathrm{num}}(x_i,y_j,\rho_k)\right|}{\max_{i,j,k}\left|\mathcal{G}f^{\mathrm{ex}}(x_i,y_j,\rho_k)\right|}\;,
\end{equation}
where $(x_i,y_j,\rho_k)$ is a grid point in $[-3,3]^2\times[0,1.55]$.

We have calculated and plotted (on a log-scale) $\epsilon_{\mathrm{abs}}$ for a wide range of sampling rates on each grid. In Figure \ref{fig:fourierfixedreal} through Figure \ref{fig:fourierfixedgyro},
each pair of side-by-side figures shows $\epsilon_{\mathrm{abs}}$ for a \textit{fixed} sampling rate of a particular grid. In Figures \ref{fig:fourierfixedreal}-\ref{fig:gyrofixedreal} we see
$\epsilon_{\mathrm{abs}}$ for a fixed sampling rate on the real space grid, in Figures \ref{fig:realfixedfourier}-\ref{fig:gyrofixedfourier} we see
$\epsilon_{\mathrm{abs}}$ for a fixed sampling rate on the Fourier space grid, and in Figures \ref{fig:realfixedgyro}-\ref{fig:fourierfixedgyro} we see
$\epsilon_{\mathrm{abs}}$ for a fixed sampling rate on the gyroradius grid. The fixed sampling rates in each case are those explained and suggested previously. The figures clearly show that with
the combined suggested choices $L_N=12$, $L_{\hat{N}}=4.5$, and $L_{N_{\rho}}=35$, the error $\epsilon_{\mathrm{abs}}$ is on the order of about $10^{-13}$. Generally, the error
decreases exponentially as we increase all three sampling rates uniformly, which is a hallmark of numerical schemes based on spectral methods.

\subsection{Run time results}\label{sec:run_time}
We are also interested in the CPU run time of our code, in particular, the relative run times of the various elements involved (i.e., forward Fourier transform, Hankel transform, Bessel function
multiplier, and inverse Fourier transform). For this purpose, we have run multiple simulations for selected sampling rates for computing the gyroaverage of the function in (\ref{eq:exactf}),
and we have collected our results in table \ref{tab:runtimes}. For each resolution, the left-hand column corresponds to the time elapsed in seconds, and the right-hand column to the time elapsed as a
percentage of the total time.

We have separated the run time data for the specific code elements from the run time data for computing the various libraries in our code (e.g., Hankel transform weights, Chebyshev grid in Fourier space, etc.).
The intended purpose of our gyroaveraging code is its use in solving the gyrokinetic-Poisson equations. Thus the gyroaveraging code is run at each time step, and so the pre-computed libraries
need be computed only once and not at each time step. Thus when analyzing the relative time of the code elements, it is not appropriate to include the time taken to build those libraries.
Hence the percentage times reported in table \ref{tab:runtimes} are percentages relative to the total time elapsed excluding the time taken to build the libraries.

Although we show the absolute run time data, given in seconds, for completeness, its importance and relevance are limited. Indeed, we have implemented our scheme in MATLAB, which is an interpreted language and therefore slower than compiled languages such as Fortran and C for the algorithms we rely on, often by an order of magnitude. Furthermore, our results were obtained by running our code on a single core, and we have not yet explored methods for accelerating our schemes through parallelization.

In contrast, the conclusions we can make from the relative elapsed times of each code element are instructive and as follows. First, we see that the relative elapsed times are approximately the same for all resolutions. Second, the Hankel
transform and Bessel function multiplication together take less than 1\% of the total time. This observation is consistent with the fact observed in section \ref{scheme} that the complexity of the Hankel transform
(i.e., $O(N_{\rho}\hat{P})$) is less than the complexity of the Fourier transforms (i.e., $O(N_{\rho}(P+\hat{P})\log(P+\hat{P}))$). The elapsed time is dominated by the time taken by the Fourier transform
and the inverse Fourier transform, with the former taking about twice as long as the latter. This difference in time is due to the use of the FST in the forward transform but not in the inverse
transform. Finally, we simply note that the computation of the libraries takes about 16\% as long as one iteration of the gyroaveraging code. 

\begin{table}
\centering
\begin{tabular}{@{}lcrrcrrcrr@{}}
\toprule
\textit{Code element} &\phantom{abc}& \multicolumn{2}{c}{\textit{Low resolution}}&\phantom{abc}& \multicolumn{2}{c}{\textit{Medium resolution}}&\phantom{abc}&\multicolumn{2}{c}{\textit{High resolution}} \\
\hline
\textit{Pre-computed libraries}          && $0.215$& 16.4\%&& $0.792$& 15.8\%&& $2.000$& 15.2\% \\
&& & && & && &\\
\textit{Fourier transform}             && $0.835$& 63.8\%&& $3.223$& 64.6\%&& $8.789$& 66.7\% \\
\textit{Hankel transform}              && $0.007$& 0.51\%&& $0.027$& 0.53\%&& $0.070$& 0.53\% \\
\textit{Bessel multiplier}             && $0.004$& 0.34\%&& $0.020$& 0.41\%&& $0.070$& 0.40\% \\
\textit{Inverse Fourier transform}     && $0.463$& 35.4\%&& $1.723$& 34.5\%&& $4.266$& 32.4\% \\
&& & && & && &\\
\textit{Total time (excl. libraries)}    && $1.309$& 100\% && $4.995$& 100\% &&$13.177$& 100\% \\
\textit{Accuracy}                      &&\multicolumn{2}{c}{$7.3\times 10^{-2}$} &&\multicolumn{2}{c}{$4.6\times 10^{-8}$}&&\multicolumn{2}{c}{$1.2\times 10^{-13}$} \\
\bottomrule
\end{tabular}
\caption{For each set of resolutions and for each code element, the CPU run time is reported in two ways: the time elapsed in seconds and the time elapsed as a
percentage of the total time (excluding the time required to pre-compute the necessary libraries). The specific sampling rates used for each resolution were
$(L_N, L_{\hat{N}},L_{N_{\rho}})=(4,2.5,11)$ for ``low resolution'', $(L_N, L_{\hat{N}},L_{N_{\rho}})=(8,3.5,23)$ for ``medium resolution'', and
$(L_N, L_{\hat{N}},L_{N_{\rho}})=(12,4.5,35)$ for ``high resolution''. The ``high resolution'' sampling rates are those sampling rates we have suggested in section \ref{scheme}.
These data were obtained by averaging the results of 1000 separate simulations of our code
in \textsc{Matlab} on a home desktop computer operating on an Intel i7-6700k processor at 4.0 GHz.}
\label{tab:runtimes}
\end{table}

\section{Summary and Discussion}\label{conclusion}
Focusing on a simple two-dimensional geometry with a uniform background magnetic field, we have proposed a new numerical scheme for the fast and high-order accurate computation of the gyroaveraged electrostatic potential for the challenging situation in which periodic boundary conditions may not be assumed. Our numerical scheme is based on a reformulation of the gyrokinetic Vlasov-Poisson system in which the electrostatic potential $\Phi$ is replaced with a new potential like function $\chi$ which also solves a Poisson equation but depends on the gyroradius variable $\rho$. The key advantage of this approach is that the function to gyroaverage has compact support. The disadvantage of our approach is that Poisson's equation has to be solved as many times as the number of discretisation points used for the $\rho$-grid. Since fast Poisson solvers are readily available, and the field solve part of gyrokinetic codes is rarely the main driver of the overall computational cost, this is a reasonable price to pay. 

Since the function to gyroaverage has compact support, we can express this gyroaverage as the composition of a Fourier transform, a Hankel transform, a multiplication by the Bessel function $J_{0}$, and an inverse Fourier transform, and each operation is numerically well defined. We perform each of these steps using spectrally accurate numerical methods with near optimal run time complexity. The total run time of our scheme is dominated by the forward and inverse Fourier transforms, so the run time complexity of our algorithm can be said to be $O(N_{\rho}(P+\hat{P})\log(P+\hat{P}))$, where $P$ is the number of spatial grid points, $\hat{P}$ the number of grid points in Fourier space, and $N_{\rho}$ the number of grid points in velocity space. By focusing on examples for which the gyroaverage can be evalualed analytically, we demonstrate that our scheme leads to geometric convergence of the numerical error for gyroaveraging, as expected. For reasonable sampling rates in real space, Fourier space, and in velocity space, we obtain an accuracy of the order of $10^{-13}$, close to round off error.

We have successfully applied our scheme to gyrokinetic-Poisson simulations of the dynamics of a non-neutral beam of particles in the median plane of a cyclotron, and verified its high accuracy in this setting \citep{GuadagniThesis}. We nevertheless see several directions for further improvement. The scheme may for example be sped up further through parallelization, and by better optimizing the size of the Fourier domain and the grid resolutions $L_{\hat{N}}$ and $L_{N_{\rho}}$. The latter could be done either through analytic formulae giving tighter estimates, or through automated computations at regularly spaced time steps. One may also wonder if there is an equivalent formulation of our scheme for electromagnetic equations, such as the gyrokinetic Vlasov-Maxwell equations for example. This is a central question for the applicability of our scheme to gyrokinetic simulations for tokamaks, and to beam dynamics studies in which relativistic effects play a significant role. These improvements are the subject of ongoing work, with progress to be reported at a later date. 

\section*{Acknowledgments}
The authors would like to thank Felix Parra for helpful discussions and the referees for valuable suggestions for improving the original manuscript.
 
\bibliographystyle{jpp}


\end{document}